\documentclass[twoside,fleqn]{article}
\usepackage{espcrc2}
\usepackage{graphicx}
\usepackage{amsbsy}
\title{Dynamical susceptibilities in a strong coupling approach}
\author{Andrij M. Shvaika\address{Institute for Condensed Matter Physics
Nat. Acad. Sci. Ukr., 1 Svientsitskii Str., Lviv UA-79011,
Ukraine}}

\begin{document}

%
%
\begin{abstract}
A general scheme to calculate dynamical susceptibilities
of strongly correlated electron systems within the dynamical
mean field theory is developed. The approach is based on an
expansion over electron hopping around the atomic limit
(within the diagrammatic technique for site operators:
projection and Hubbard ones) in infinite dimensions.
As an example, the Falicov-Kimball and simplified
pseudospin-electron models are considered and
analytical expressions for the dynamical susceptibilities are
obtained.
\end{abstract}

\maketitle

In the last decade the main achievements in the theory of the strongly
correlated electron systems are connected with the development of the
Dynamical Mean Field Theory (DMFT) which is exact in the $d=\infty$
limit \cite{Kotliar}. It was shown by Metzner and Vollhardt
\cite{MetznerVoll,Metzner} that in the $d=\infty$ limit
self-energies are single-site quantities (do not depend on wave
vector) which leads to a significant simplification.
The same is true for the
four-vertices in the Bethe-Salpeter equation for susceptibilities
(see \cite{Kotliar}). 

The aim of this article is to develop a general scheme to calculate
dynamical susceptibilities within a diagrammatic technique for site
operators (projection or Hubbard ones) for strongly correlated
electron systems described by the general statistical operator
\begin{equation}
\hat{\rho} = e^{-\beta \hat{H}_{0}}\times
\label{eq1}
\end{equation}
\[
T\!\exp \left\{\!-\!\int\limits_{0}^{\beta}\!\! d\tau\!
\int\limits_{0}^{\beta}\!\!d\tau' \sum\limits_{ij\sigma} t_{ij}^{\sigma}
(\tau-\tau') a_{i\sigma}^{\dag}(\tau) a_{j\sigma}(\tau') \!\right\},
\]
where $\hat{H}_{0} = \sum\limits_{i} \hat{H}_{i}$ is a sum of the single-site 
contributions. Our approach is based on an expansion over electron hopping
around the atomic limit \cite{Metzner} (see also \cite{Shvaika}) instead of
an expansion in the local interaction \cite{Kotliar}.

In the $d=\infty$ limit, the lattice problem with
$t_{ij}^{\sigma}(\tau-\tau')=\delta(\tau-\tau')t_{ij}/\sqrt{d}$,
is mapped onto an effective atomic problem with a dynamical mean
field $t_{ij}^{\sigma}(\tau-\tau')=\delta_{ij}
J_{\sigma}(\tau-\tau')$.
Single-ele\-c\-t\-ron Green's functions for the lattice
$G_{\sigma}(\omega_{\nu},\boldsymbol{k})=
\left[\Xi_{\sigma}^{-1}(\omega_{\nu})-t_{\boldsymbol{k}}\right]^{-1}$
and the effective atomic problem $G_{\sigma}^{(a)}(\omega_{\nu})=
\left[\Xi_{\sigma}^{-1}(\omega_{\nu})-J_{\sigma}(\omega_{\nu})\right]^{-1}$
are determined by the same single-site irreducible parts
$\Xi_{\sigma}(\omega_{\nu})$ \cite{Metzner} and a closed set of
equations for $\Xi_{\sigma}(\omega_{\nu})$ and $J_{\sigma}(\omega_{\nu})$
can be written \cite{Kotliar}. Here, irreducible parts are 
contributions to the single-site Green's function which cannot be
divided into parts by cutting one hopping line.

In a similar way, the expansion for correlation functions built on
operators $\hat A$ and $\hat B$ in terms of the hopping is
\begin{equation}\label{L-suscept}
\langle T\!\hat A(\tau)\hat B(\tau')\rangle=
\raisebox{-53pt}{\includegraphics[width=45mm]{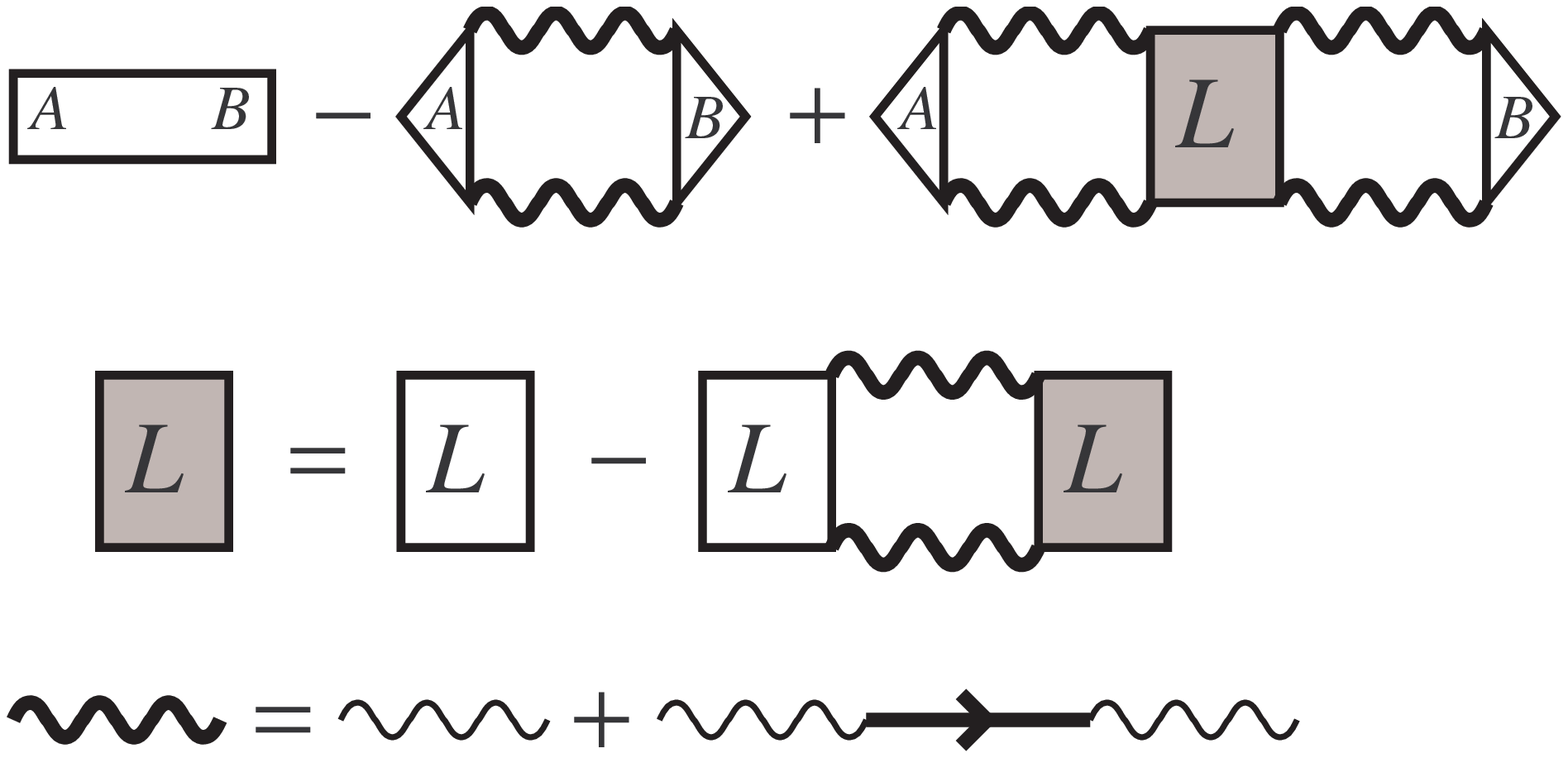}}
\end{equation}
where thin wavy lines denote electron hopping integrals
$t_{ij}^{\sigma}(\tau-\tau')$ and arrows denote single-particle
Green's functions. In eq. (\ref{L-suscept}), \frame{\phantom{OOO}},
$\Bigl\langle\hspace{-.28em}\Bigr|$ and \framebox{\em L}
are single-site quantities, which are the same for the lattice and the effective
atomic problems and
are generalized many-particle Green's functions \cite{Shvaika} which will be
calculated within DMFT.

To do this, we calculate a two-particle Green's function for the
effective atomic problem
\[
\langle T a^\dag_1a_2a^\dag_3a_4\rangle_{(a)}\equiv
\raisebox{-8pt}{\includegraphics[width=8mm]{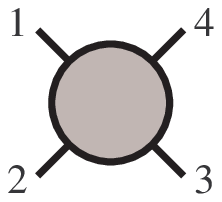}},
\]
which, on the other hand, can be written in the following way:
\[
\raisebox{-8pt}{\includegraphics[width=73mm]{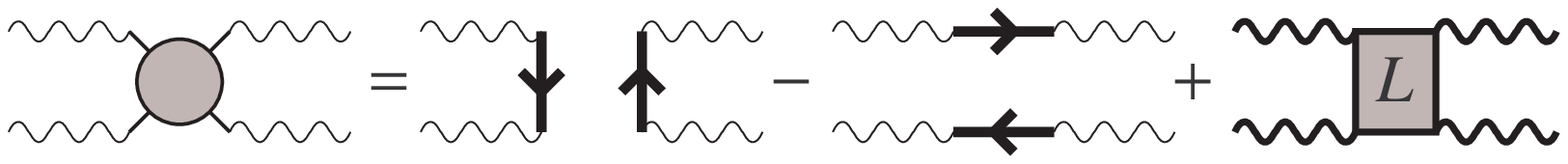}}.
\]
Combining this equation with the equation for the four-vertex of the
atomic problem allows us
to calculate the single-site four-vertex \framebox{\em L}. Similarly,
we can calculate other single-site vertices
\frame{\phantom{OOO}} and $\Bigl\langle\hspace{-.28em}\Bigr|$ and,
finally, determine the correlation functions (\ref{L-suscept}) for the
lattice.

As an example, we consider a binary alloy model
\begin{equation}
H_i=\frac g2 P^+_i n_i - \frac g2 P^-_i n_i - \frac h2 (P^+_i -
P^-_i) -\mu n_i,
\label{binall}
\end{equation}
where $P^\pm_i = \frac 12 \pm S^z_i$ for the $U=0$ pseudospin-electron (PE)
model, $P_i^+=c_i$, $P_i^-=1-c_i$ for a binary alloy, and $a_{i\sigma}\to d_i$,
$n_i=d^\dag_i d_i$,
$P_i^+=f^{\dag}_i f_i$, $P_i^-=1-f^{\dag}_i f_i$ for the Falicov--Kimball
(FK) model \cite{Falicov}.
The single-particle Green's function for the effective atomic problem
is a coherent sum of the Green's functions for subspaces
$S_i^z={\pm}\frac12$ and is equal to \cite{FK,JPS}
\[
  G_{\sigma}^{(a)}(\omega_{\nu}) =
  \frac{\left\langle P^+\right\rangle}
  {i\omega_{\nu}+\mu-J_{\sigma}(\omega_{\nu})-\frac g2}
\]
\[
  \qquad\qquad+\frac{\left\langle P^-\right\rangle}
  {i\omega_{\nu}+\mu-J_{\sigma}(\omega_{\nu})+\frac g2}
\]
which allows us to find solutions for $\Xi_{\sigma}(\omega_{\nu})$ and
$J_{\sigma}(\omega_{\nu})$.

For the model (\ref{binall}) all many-particle Green's functions for
the effective atomic problem can be obtained in an analytical form
which allows us to calculate dynamical susceptibilities: pseudospin
\[
\chi^{S^zS^z}(\omega_m,\boldsymbol{q})=
\frac{\delta(\omega_m)\Delta^2_{S^z}}{T-\Theta(T,\boldsymbol{q})},
\]
charge
\[
\chi^{nn}(\omega_m,\boldsymbol{q})=
\frac{\delta(\omega_m)\Delta^2_n}{T-\Theta(T,\boldsymbol{q})} + K^{nn}(\omega_m,\boldsymbol{q})
\]
and mixed
\[
\chi^{nS^z}(\omega_m,\boldsymbol{q})=\chi^{S^zn}(\omega_m,\boldsymbol{q})=
\frac{\delta(\omega_m)\Delta_n\Delta_{S^z}}{T-\Theta(T,\boldsymbol{q})},
\]
where
\[
\Theta(T,\boldsymbol{q})=\frac1\beta\sum_{\nu\sigma}
\frac{\Lambda^2_{\sigma \nu}
(\chi_{\sigma \nu0}(\boldsymbol{q})-\tilde\chi_{\sigma \nu})}
{\tilde\chi^{2}_{\sigma \nu0}+\Lambda^2_{\sigma \nu}
(\chi_{\sigma \nu0}(\boldsymbol{q})-\tilde\chi_{\sigma \nu0})},
\]
\[
K^{nn}(\omega_m,\boldsymbol{q})=
\]
\[
\frac1\beta\sum_{\nu\sigma}
\frac{\tilde\chi_{\sigma \nu m}(\tilde\chi_{\sigma \nu m}-
\Lambda_{\sigma \nu}\Lambda_{\sigma \nu+m})
\chi_{\sigma \nu m}(\boldsymbol{q})}
{\tilde\chi^2_{\sigma \nu m}+
\Lambda_{\sigma \nu}\Lambda_{\sigma \nu+m}
(\chi_{\sigma \nu m}(\boldsymbol{q})-\tilde\chi_{\sigma \nu m})},
\]
\[
\Lambda_{\sigma \nu}=\frac{g\sqrt{\langle P^+\rangle\langle P^-\rangle}}
{(i\omega_{\nu}+\mu-J_{\sigma}(\omega_{\nu}))^2-\frac{g^2}4},
\]
\[
\Delta_n=\frac1\beta\sum_{\nu \sigma}
\frac{\Lambda_{\sigma \nu}\tilde\chi_{\sigma \nu0}
\chi_{\sigma \nu0}(\boldsymbol{q})}
{\tilde\chi^2_{\sigma \nu0}+\Lambda^2_{\sigma \nu}
(\chi_{\sigma \nu0}(\boldsymbol{q})-\tilde\chi_{\sigma \nu0})},
\]
\[
\Delta_{S^z}=\sqrt{\langle P^+\rangle\langle P^-\rangle},
\]
\[
\chi_{\sigma \nu m}(\boldsymbol{q})=
-\frac1N\sum_{\boldsymbol{k}}G_\sigma(\omega_{\nu},\boldsymbol{k})
G_\sigma(\omega_{\nu+m},\boldsymbol{k}+\boldsymbol{q}),
\]
\[
\tilde\chi_{\sigma \nu m}=-G_\sigma(\omega_{\nu})G_\sigma(\omega_{\nu+m}).
\]
The expression for $\Theta(T,\boldsymbol{q})$ coincides with the one obtained by
Freericks \cite{Freericks} and it is known that
the ground state of the model (\ref{binall}), for a fixed average value of the
pseudospin, is not uniform and shows either
commensurate order, incommensurate order, or phase separation
\cite{Freericks}. On the other hand, in the case of a fixed
value of the field $h$ the possibility of a
uniform first-order phase transition (bistability) appears \cite{JPS,Chung}.


\begin{thebibliography}{99}
\bibitem{Kotliar} A.~Georges, G.~Kotliar, W.~Krauth, M.J.~Rosenberg,
        Rev. Mod. Phys. 68 (1996) 13.
\bibitem{MetznerVoll} W.~Metzner, D.~Vollhardt, Phys. Rev. Lett.
        62 (1989) 324.
\bibitem{Metzner} W.~Metzner, Phys. Rev. B 43 (1991) 8549.
\bibitem{Shvaika} A.M.~Shvaika, cond-mat/9911255.
\bibitem{Falicov} L.M.~Falicov, J.C.~Kimball, Phys. Rev. Lett. 22 (1969)
997.
\bibitem{FK} U.~Brandt, C.~Mielsch, Z. Phys. B 75 (1989) 365;
        79 (1990) 295; 82 (1991) 37.
\bibitem{JPS} I.V.~Stasyuk, A.M.~Shvaika, J. Phys. Studies
         3 (1999) 177.
\bibitem{Freericks} J.K.~Freericks, Phys. Rev. B 47
        (1993) 9263.
\bibitem{Chung} W.~Chung, J.K.~Freericks, Phys. Rev. B 57 (1998) 11955.
\end{thebibliography}
\end{document}